\def\gsim{\:\raisebox{-0.5ex}{$\stackrel{\textstyle>}{\sim}$}\:}

\documentstyle[sprocl,psfig]{article}

\def\Journal#1#2#3#4{{#1} {\bf #2}, #3 (#4)}

\def\NPB{{\em Nucl. Phys.} B}
\def\PLB{{\em Phys. Lett.}  B}
\def\PRL{\em Phys. Rev. Lett.}
\def\PRD{{\em Phys. Rev.} D}

\def\APJ{\em Astrophys. J} 

\def\be{\begin{equation}}
\def\ee{\end{equation}}
\def\bea{\begin{eqnarray}}
\def\eea{\end{eqnarray}}

\begin{document}
\title{\vglue -3.2cm 
\hfill{\small hep-ph/9802295}\\
\hfill{\small FTUV/98-15}\\
\hfill{\small IFIC/98-15}
\vskip 1.2cm
Supernova Bounds on Neutrino Properties: a mini-review\footnote{Talk given 
in Valencia 97: International Workshop on 
"Physics Beyond the Standard Model: from Theory to Experiment" 
Oct.  13-17, 1997, Valencia, Spain
}}

\author{Hiroshi Nunokawa\footnote{Present address: 
Inst. de F\'{\i}sica Gleb Wataghin, 
Univ. Estadual de Campinas - UNICAMP, 
P.O. Box 6165, 13083-970 Campinas SP BRAZIL, 
E-mail: nunokawa@ifi.unicamp.br
}}

\address{Instituto de F\'{\i}sica Corpuscular - C.S.I.C., 
Departament de F\'{\i}sica Te\`orica, Universitat de Val\`encia,
46100 Burjassot, Val\`encia, SPAIN}


\maketitle\abstracts{This talk summarizes our recent work 
which studied the impact of resonant neutrino conversion induced 
by some non-standard neutrino properties beyond mass and mixing, 
such as neutrino magnetic moment, lepton-flavor non-universality 
as well as flavor changing neutral current interactions
in SUSY models with broken $R$ parity, on supernova physics. 
}

\section{Introduction}
Neutrino flavor conversion could cause some significant 
influence on supernova physics \cite{review}. 
In this talk we discuss the effect of such conversion 
induced by non-standard properties of neutrinos, not just by 
mass and mixing, on supernova physics. 
In particular, we consider the effect on 
neutrino shock-reheating, supernova heavy elements nucleosynthesis 
as well as $\bar{\nu}_e$ signal, and show that in some 
case rather stringent limits on model
parameters can be obtained. 
\subsection{Some basic features of supernova neutrinos}
A type-II  supernova occurs when a massive star 
($M \gsim 8M_\odot$) has reached the last stage of 
its life \cite{suzuki}. 
Almost all of the gravitational binding energy 
of the final neutron star (about $\sim 10^{53}$ erg) is 
radiated away in form of neutrinos. 
The individual neutrino luminosities in supernovae 
are approximately the same but the individual neutrino 
energy distributions are very different
because they interact differently with the star material, 
as following reactions show, 
\begin{eqnarray}
\label{nu-n}
\nu_e+n&\to & p+e^-,\\
\label{nu-p}
\bar\nu_e+p& \to & n +e^+,\\
\label{nu-N}
\nu +  N & \to &\nu +  N, \ \ (N=p,n).
\end{eqnarray}
Since the cross sections of the charged-current reaction 
is larger than that of the neutral-current one and 
there are more neutrons than protons, 
the $\nu_e$'s have the largest interaction rates with 
the matter and hence thermally decouple at the lowest 
temperature. 
On the other hand, 
$\nu_{\tau(\mu)}$ and $\bar\nu_{\tau(\mu)}$'s lack 
the the charged-current absorption reactions 
on the free nucleons 
inside the neutron star and hence thermally decouple at the 
highest temperature. 
As a result, the average neutrino energies satisfy 
the following hierarchy:
\begin{equation}
\label{hierarchy}
\langle E_{\nu_e}\rangle <\langle E_{\bar\nu_e}\rangle <
\langle E_{\nu_{\tau(\mu)}}\rangle 
\approx\langle E_{\bar\nu_{\tau(\mu)}}\rangle.
\end{equation}
Typically, the average supernova neutrino energies are, 
$\langle E_{\nu_e}\rangle \approx 11\ \mbox{MeV},\ 
\langle E_{\bar\nu_e}\rangle
\approx 16\ \mbox{MeV},\ \langle E_{\nu_{\tau(\mu)}}\rangle
\approx \langle
E_{\bar\nu_{\tau(\mu)}}\rangle\approx 25\ \mbox{MeV}$. 
\subsection{Impact of neutrino oscillation on supernova physics}
Here we very briefly review some significant effects 
of neutrino oscillation, which occurs if
neutrinos are massive and mixed, 
on supernova physics, 
studied in some previous work. 
First issue is concerned with the neutrino conversion effect 
on shock re-heating in the delayed explosion scenario \cite{delayed}. 
If neutrinos are massive and mixed and follow the mass hierarchy 
as observed in the quark sector, one expects MSW resonant 
conversion \cite{MSW} between $\nu_e$ and $\nu_\mu$ or  $\nu_\tau$ 
inside supernova. 
If the conversion occurs between the neutrinosphere and 
the stalled shock this can help the explosion \cite{fuller}. 
Due to the conversion the energy spectra of 
$\nu_e$ and $\nu_\mu$ or  $\nu_\tau$ can be swapped and
hence $\nu_e$ would have larger average energy leading 
to a larger energy deposition by reactions in 
eqs. (\ref{nu-n}) and (\ref{nu-p}) so that the stalled 
shock would be re-energized. 

Second issue is the impact on heavy elements nucleosynthesis 
in supernova. 
To have successful $r$-process the site must be neutron 
rich, i.e. $Y_e < 0.5$ where $Y_e$ is number of 
electron per baryon. The $Y_e$ value is mainly 
determined by the competition between the 
two absorption reactions in eqs. (\ref{nu-n}) and (\ref{nu-p}). 
In the standard supernova model the latter process 
is favoured due to the higher average energy of 
$\bar\nu_e$ which guarantees the neutron richness. 
If the neutrino oscillations do occur between the 
neutrinosphere and the region relevant for $r$-process
the site can be driven to proton-rich due to the 
reaction (\ref{nu-n}) and therefore, $r$-process could
be prevented \cite{qian}. 

Third argument is the effect of neutrino oscillation 
on $\bar{\nu}_e$ signal in the terrestrial detector. 
It is discussed that from SN1987A data \cite{kamimb} 
the large oscillation between $\bar{\nu}_e$ and 
$\bar{\nu}_\mu$ or $\bar{\nu}_\tau$ 
is disfavoured since the oscillation 
can induce harder $\bar{\nu}_e$ spectra than 
the observed one \cite{ssb}. 

Finally, although we will not discuss this further in this 
talk, we also mention that neutrino oscillation in the 
presence of polarization of the medium in the star
due to the strong magnetic field could lead to some 
interesting consequences \cite{pol}. 
\section{Neutrino oscillation induced by non-standard neutrino properties}
Here we discuss our main interest, the effect of neutrino oscillation 
induced by some non-standard properties of neutrinos other than 
mass and mixing. 
\vglue -0.8cm 
\subsection{Resonant Spin-Flavor Precession}
First discussion is devoted to the impact of 
the resonant spin-flavor precession (RSFP) \cite{rsfp}, on 
supernova nucleosynthesis and dynamics \cite{nqf,rsfpsn}. 
Transition magnetic moment of neutrinos can cause 
simultaneous change of helicity and flavor of neutrinos \cite{sv}. 
Moreover, in matter, for the case of Majorana neutrinos, 
$\nu_\tau$ (or $\nu_\mu$) can resonantly convert into 
$\bar{\nu}_e$ and vice versa  \cite{rsfp}. 

\begin{figure}[htb]
\vskip -0.95cm
\centerline{\protect\hbox{
\psfig{file=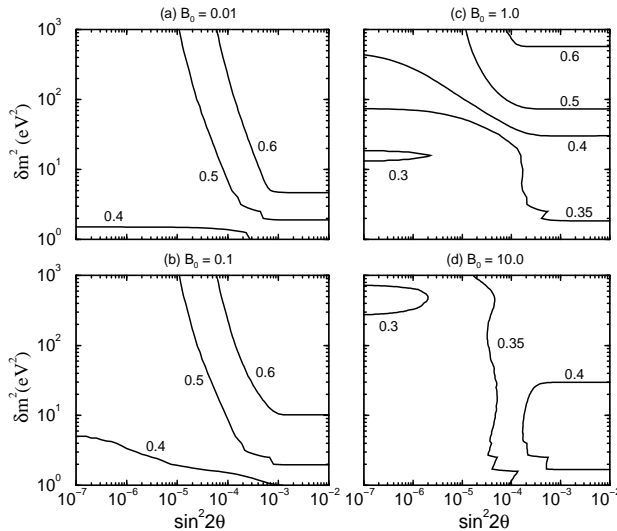,height=7.5cm,width=8.8cm,angle=-90}
}}
\vskip -0.2cm
\caption{Contour plots for the electron fraction $Y_e$ in 
the $\sin^22\theta-\delta m^2$ plane in the 
$r$-process site, for the magnetic field profile 
$B(r)=B_0[r_0/r]^2\times 10^{12}$ G
($r_0=10$ km) with $B_0$ = (a) 0.01, (b) 0.1 (c) 1.0 and (d) 10.0.
The transition magnetic moment of neutrino is assumed to be 
$10^{-12} \mu_B$ where $\mu_B$ is Bohr magneton. 
(From ref. \protect{\cite{nqf}}.)}
\label{fig:rsfpye}
\vskip -0.15cm
\end{figure}

Contrary to the case of MSW effect \cite{qian}, RSFP can 
decrease the electron fraction in the $r$-process 
region even if the MSW effect co-exist \cite{nqf}. 
This is because RSFP can swap the energy spectra of 
$\bar{\nu}_e$ and $\nu_\tau$ (or $\nu_\mu$), 
leading to the larger $\bar{\nu}_e$ energy and increasing the 
neutron richness due to the reaction (\ref{nu-p}). 
In Fig. \ref{fig:rsfpye} we plot the expected electron fraction 
in the presence of RSFP with flavor mixing. 
We see that for weaker magnetic field, the result 
agrees with Qian {\it et al.} \cite{qian} 
but for stronger magnetic field $Y_e$ can be lower than 
the expected value in the absence of oscillation ($\sim 0.4$), 
which implies the enhancement of $r$-process. 
%
\begin{figure}[htb]
\vskip -0.6cm
\centerline{\protect\hbox{
\psfig{file=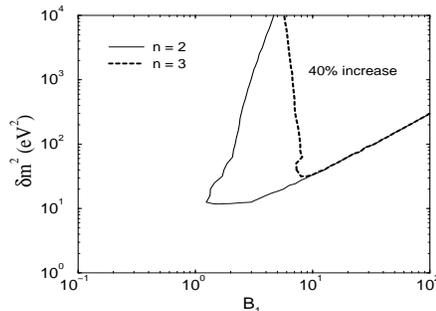,height=4.6cm,width=6.3cm,angle=-90}
}}
\vskip -0.3cm
\caption{
Regions of $B_1-\delta m^2$ parameter space
where the increase in the total reheating rate 
is 40 \%. The solid and dashed lines are for the
magnetic field profile $B(r)=B_1[r_0/r]^n \times 10^{12}$ G 
($r_1=100$ km) with $n=2$ and 3, respectively.
(From ref. \protect{\cite{nqf}}.) }
\label{fig:rsfpshock}
\vglue -0.5cm
\end{figure}
%

RSFP can help the shock reheating by neutrino since the 
energy of $\bar{\nu}_e$ is increased \cite{alps}. 
We estimate the neutrino energy deposition rate at 
the stalled shock, which is normalized to 1 
in the absence of any kind neutrino conversion. 
We present our results in Fig. \ref{fig:rsfpshock}. 
We see that for some range of parameter, the total 
reheating rate can be increased as large as 40 \%. 
We note, however, that 
RSFP conversion could be in conflict with SN1987A data
since the energy spectra of $\bar{\nu}_e$ get 
harder \cite{ssb}. 
\subsection{
Massless neutrino conversion induced by flavor non-universality}
In some case neutrinos can mix and even resonantly convert 
into another flavor in matter even if they are 
strictly massless \cite{valle}. 
Here we focus on a particular scenario of massless-neutrino 
conversion suggested in ref. \cite{valle}. 
It is possible with extra gauge singlet neutrinos
and requiring lepton number conservation to keep the 
standard neutrinos massless but mixed.
It is, however, very difficult to observe this mixing 
through conventional neutrino oscillation experiments 
because the phase can not be developed in vacuum 
since neutrinos are massless. 
However, in matter neutrinos can acquire non-trivial phase 
due to flavor non-universality. 
Here, we consider the $\nu_e - \nu_\tau $ system and we define 
the measure of non-universality as \cite{massless}
$\eta \equiv \frac{1}{2} (h_\tau^2-h_e^2)$, 
where
$h_\tau$ and $h_e$ denote the deviation from the
standard coupling which are assumed to be as 
large as a few \% especially for $h_\tau$. 
It has been shown that for nonzero $\eta$, 
the resonant neutrino conversion between 
$\nu_e - \nu_\tau $ and $\bar{\nu}_e - \bar{\nu}_\tau $ 
can occur \cite{valle}. 
Such massless neutrino conversion is different from the usual 
MSW conversion in that the conversion probability is energy 
independent and it occurs for both neutrino and anti-neutrino 
channels simultaneously. 

\begin{figure}[htb]
\vskip -0.57cm
{\hskip -0.8cm
\psfig{file=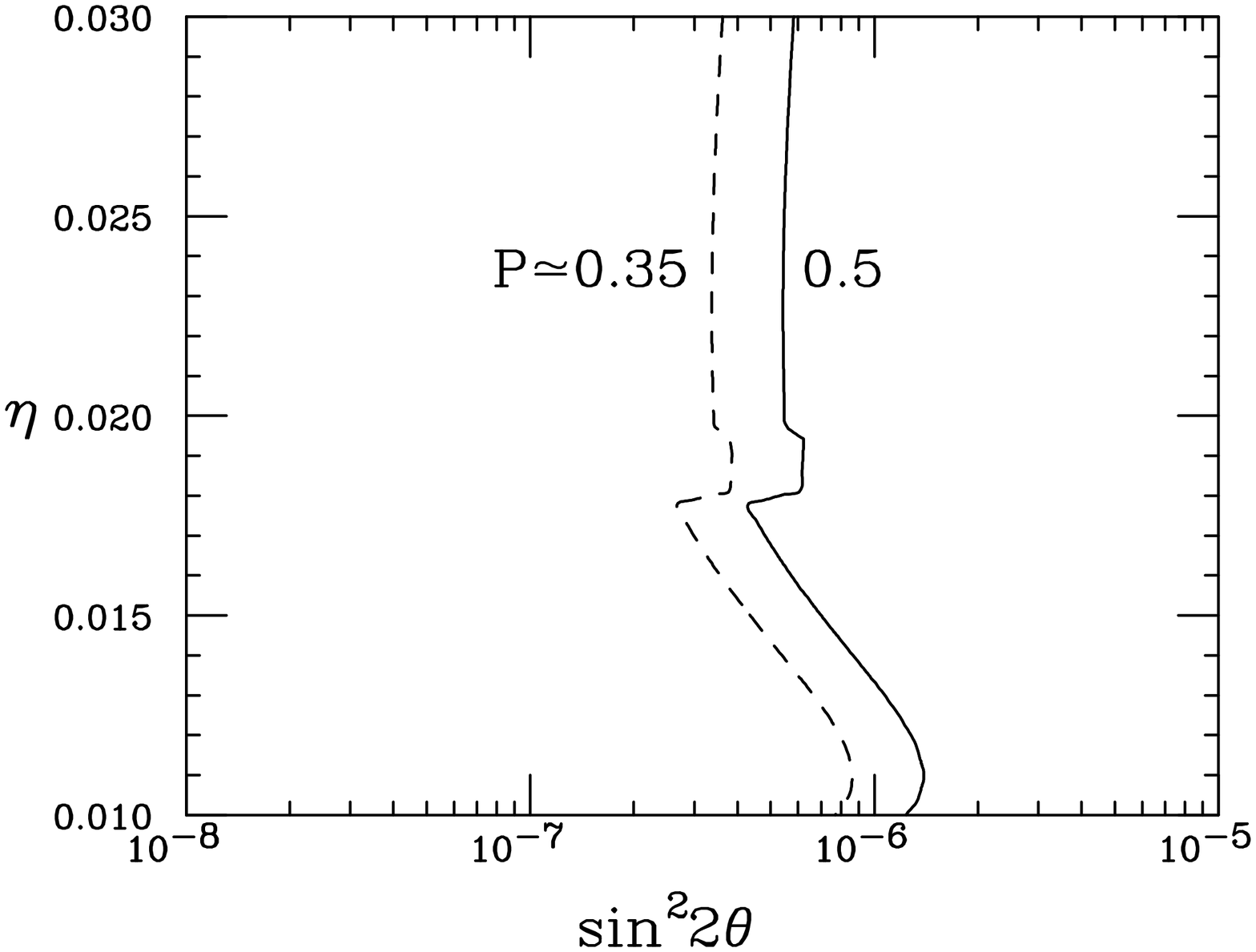,height=4.6cm,width=6.0cm,angle=90}
}
\vskip -4.595cm
{\hglue 5.4cm
\psfig{file=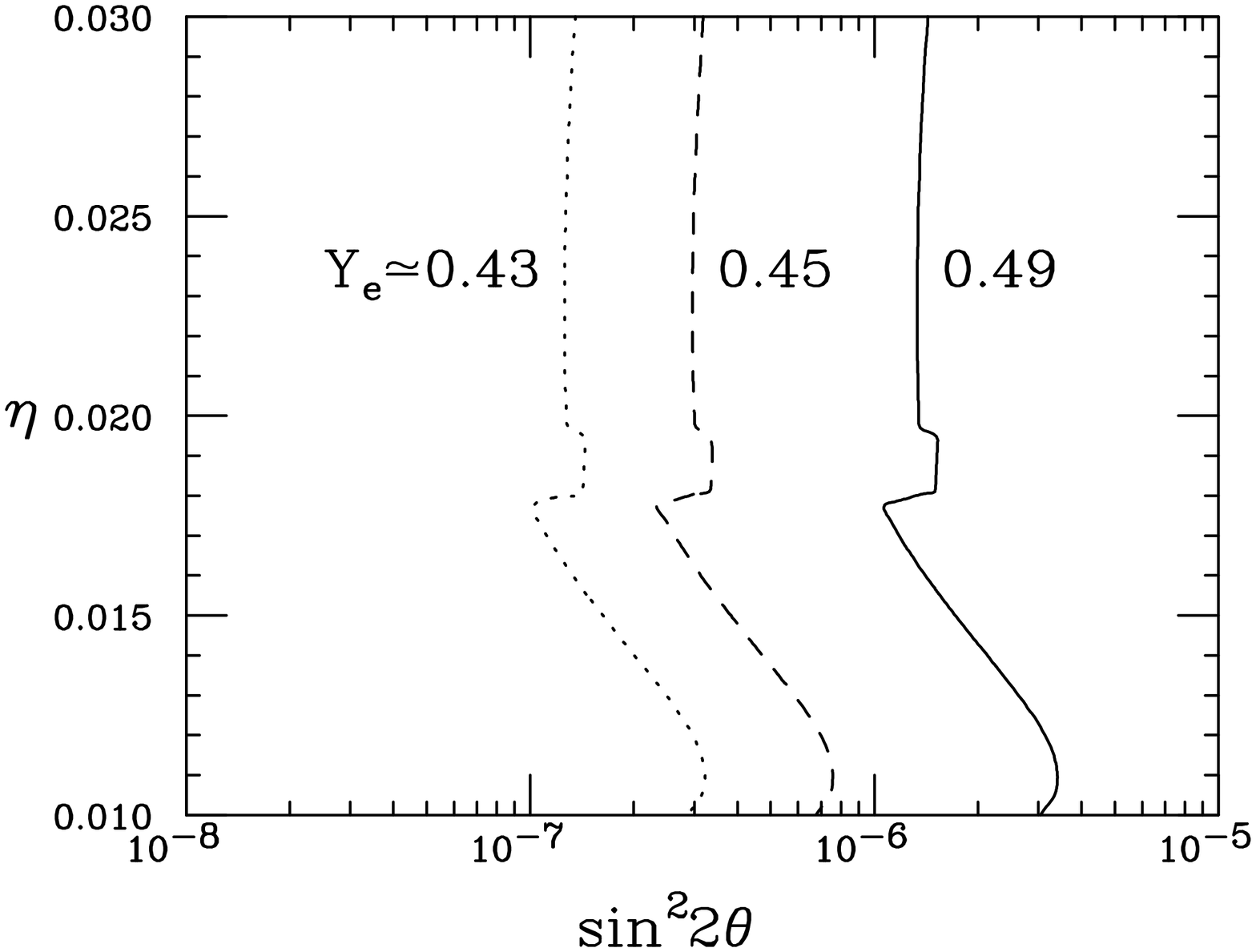,height=4.6cm,width=6.0cm,angle=90}
}
\vglue -0.6cm
\caption{
Left panel shows the constraints on massless-neutrino mixing 
from the detected SN1987A $\bar\nu_e$ energy spectra. 
The region to the right of the dashed (solid) lines are excluded by 
the detection data for an allowed conversion probability of $P<0.35$ (0.5).
Right panel is similar to the left one but from the supernova 
$r$-process nucleosynthesis. The region to the
right of the dotted, dashed and solid lines are excluded for the 
required values of $Y_e<0.43$, 0.45, and 0.49, respectively,
in the $r$-process. (From ref. \protect{\cite{massless}}.)
}
\label{fig:mless}
\vglue -0.5cm
\end{figure}

Here we argue that such conversion can be in conflict 
with SN1987A $\bar{\nu}_e$ signal \cite{ssb} and $r$-process 
scenario \cite{qian} and hence we can constrain the relevant 
parameters \cite{massless}. 
We show our results in Fig. \ref{fig:mless} (see the caption).

\subsection{FCNC induced neutrino conversion in 
SUSY models with broken $R$ parity}
It has been discussed that neutrino conversion 
can be induced by flavor changing neutral current (FCNC) 
interaction even if neutrinos are unmixed \cite{rsusy0}. 
Here, we focus on a particular conversion mechanism \cite{rsusy0} 
induced by some new interactions between neutrinos and matter 
mediated by the scalar partners of quarks and leptons in 
supersymmetric extension of the standard model with explicitly 
broken $R$-parity. 
\begin{figure}[htb]
\vskip -0.755cm
{\hskip -0.8cm
\psfig{file=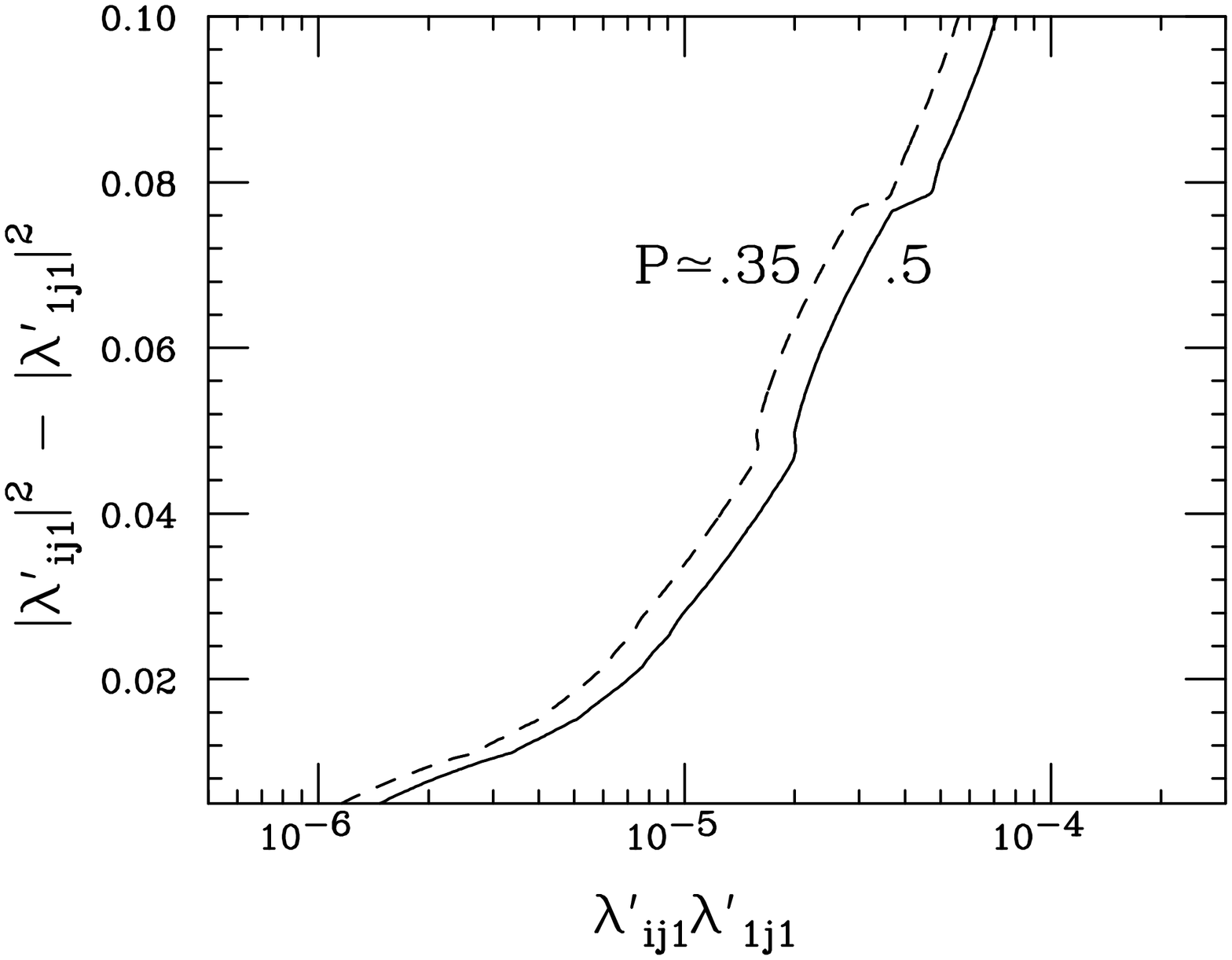,height=4.6cm,width=6.0cm,angle=90}
}
\vskip -4.65cm
{\hglue 5.4cm
\psfig{file=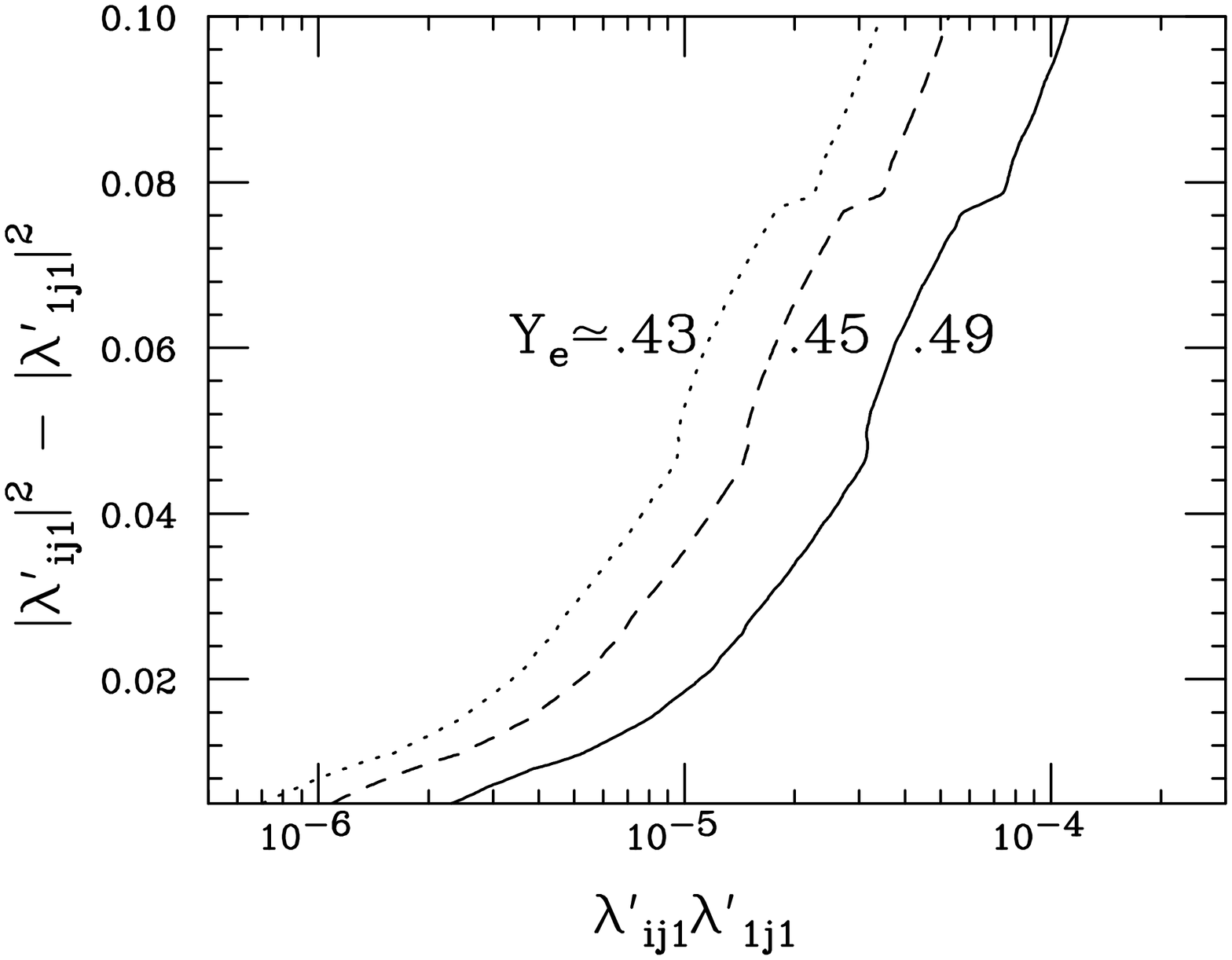,height=4.6cm,width=6.0cm,angle=90}
}
\vglue -0.55cm
\caption{Similar plots as in Fig. 3 but for different 
conversion mechanism induced by FCNC interaction 
in SUSY models with broken $R$ parity. 
(From ref. \protect{\cite{rsusy}}.)
}
\label{fig:rsusy}
\vglue -0.2cm
\end{figure}
\begin{figure}[htb]
\vskip -0.6cm
{\hskip -0.8cm
\psfig{file=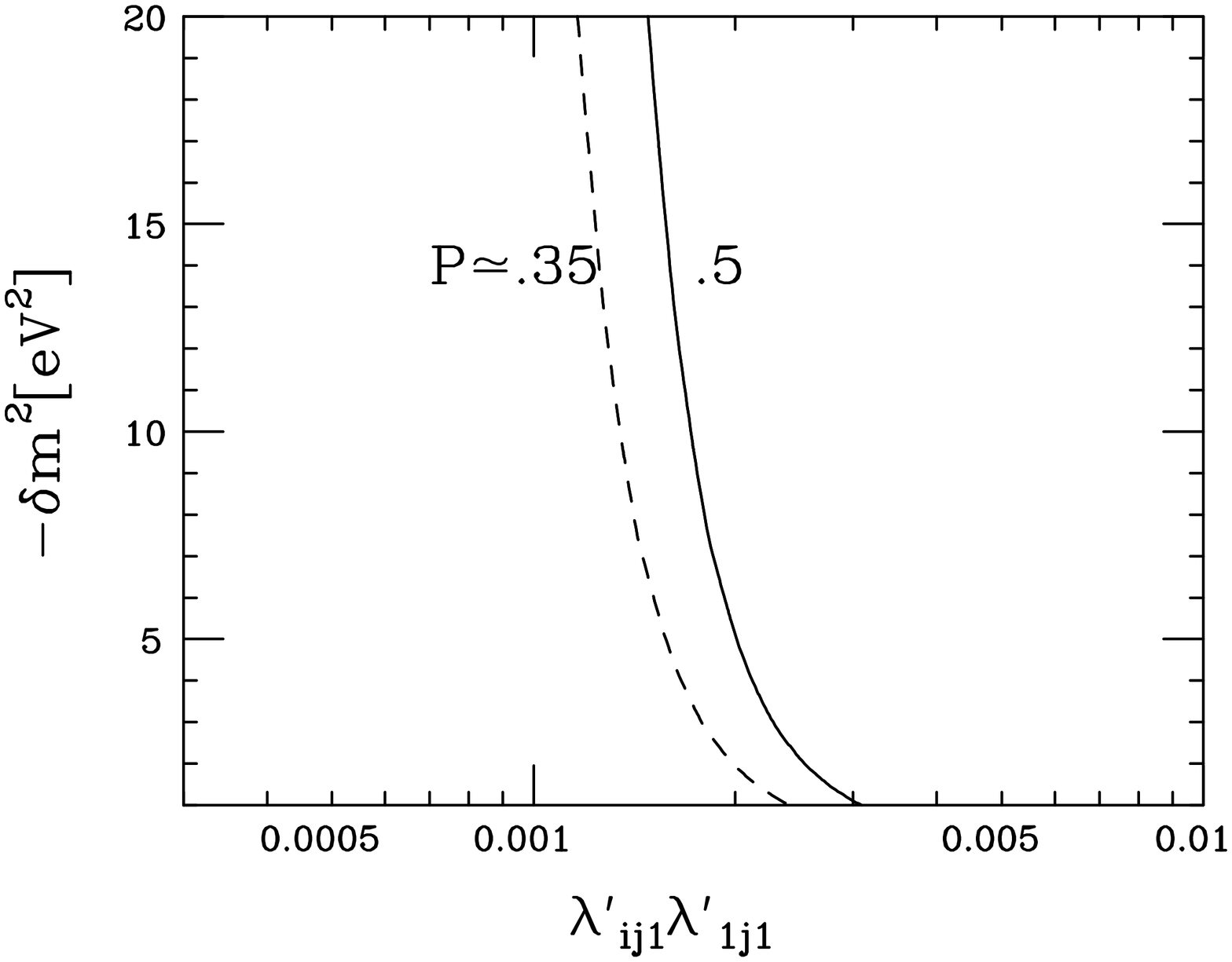,height=4.5cm,width=6.0cm,angle=90}
}
\vskip -4.57cm
{\hglue 5.4cm
\psfig{file=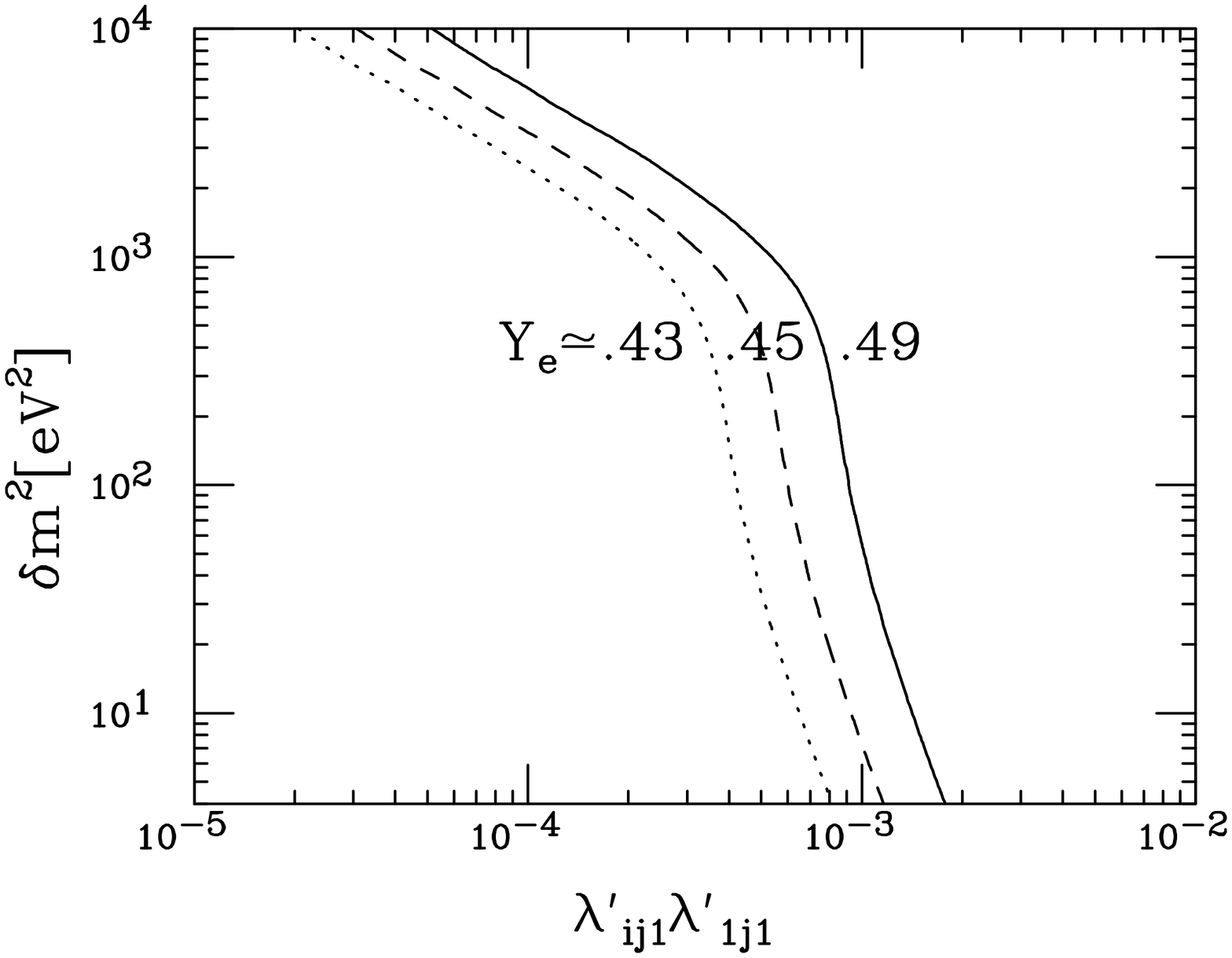,height=4.5cm,width=6.0cm,angle=90}
}
\vglue -0.55cm
\caption{Similar plot as Fig. 4 but for non-zero 
neutrino masses. We note, however, that if 
$\delta m^2 \equiv m_2^2 - m_1^2$ is negative
only the constraints from SN1987A $\bar\nu_e$ energy 
spectra (left panel) is obtained whereas $\delta m^2$ is positive 
only the constraints from 
$r$-process is obtained (right panel) is obtained. 
(From ref. \protect{\cite{rsusy}}.)}
\vglue -0.5cm
\end{figure}
In Fig. \ref{fig:rsusy} 
we show the parameter region excluded by SN1987A signal \cite{ssb} 
and by $r$-process \cite{qian} assuming the vanishing neutrino masses. 
We note that some features of neutrino conversion in this 
case is similar to the one we discussed in sec. 2.2, i.e., 
the energy independence and simultaneous conversion 
of neutrinos and anti-neutrinos. 

We also consider the case where neutrino masses are not negligible 
because neutrino masses are naturally induced in this model 
at one loop level.  
We present in Fig. 5 our results for this case. 
\subsection{Neutrino conversion into sterile state}
Finally we discussed the case where neutrinos are 
mixed with some sterile state. 
We have reanalysed the impact of resonant conversion of 
electron neutrinos into some sterile state on supernova 
physics \cite{sterile0} assuming the mass of the sterile 
state to be in the cosmologically significant range, i.e. 1-100 eV, 
the range relevant as dark matter component in the universe
\cite{kolb}. Here we consider the system of 
$\nu_e$ and $\nu_s$ (and their anti-partners ) 
with non-zero masses and mixings and neglect 
the mixing among other flavors.  
Due to non-monotonic behaviour of the potential 
above the neutrinosphere not only neutrinos but also 
anti-neutrinos can convert into sterile state \cite{sterile}.  

\begin{figure}[htb]
\vskip -0.58cm 
\centerline{\hskip -0.1cm \protect\hbox{
\psfig{file=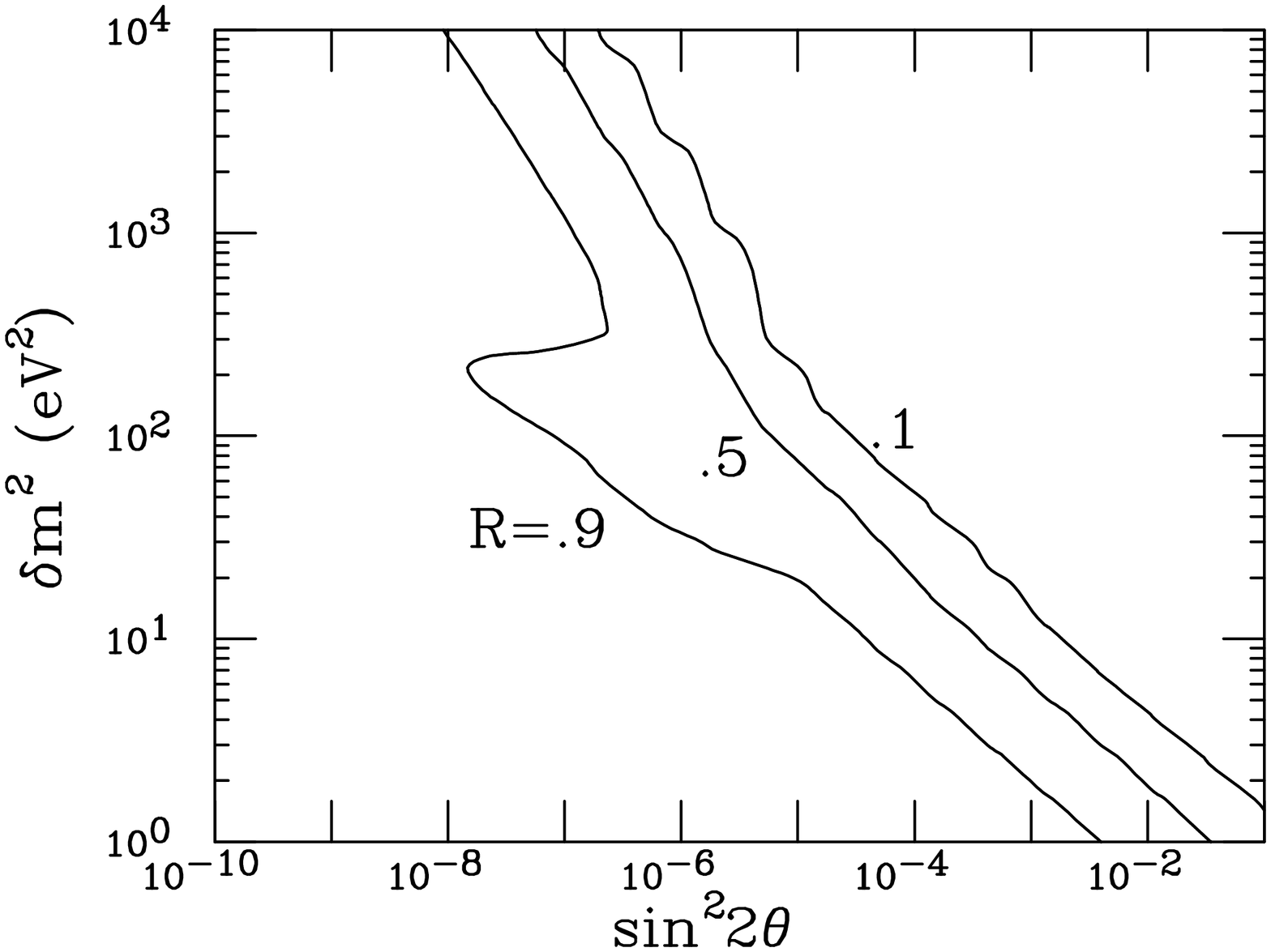,height=4.2cm,width=6.3cm,angle=90}
}}
\vskip -0.67cm
{\hskip -0.25cm
\psfig{file=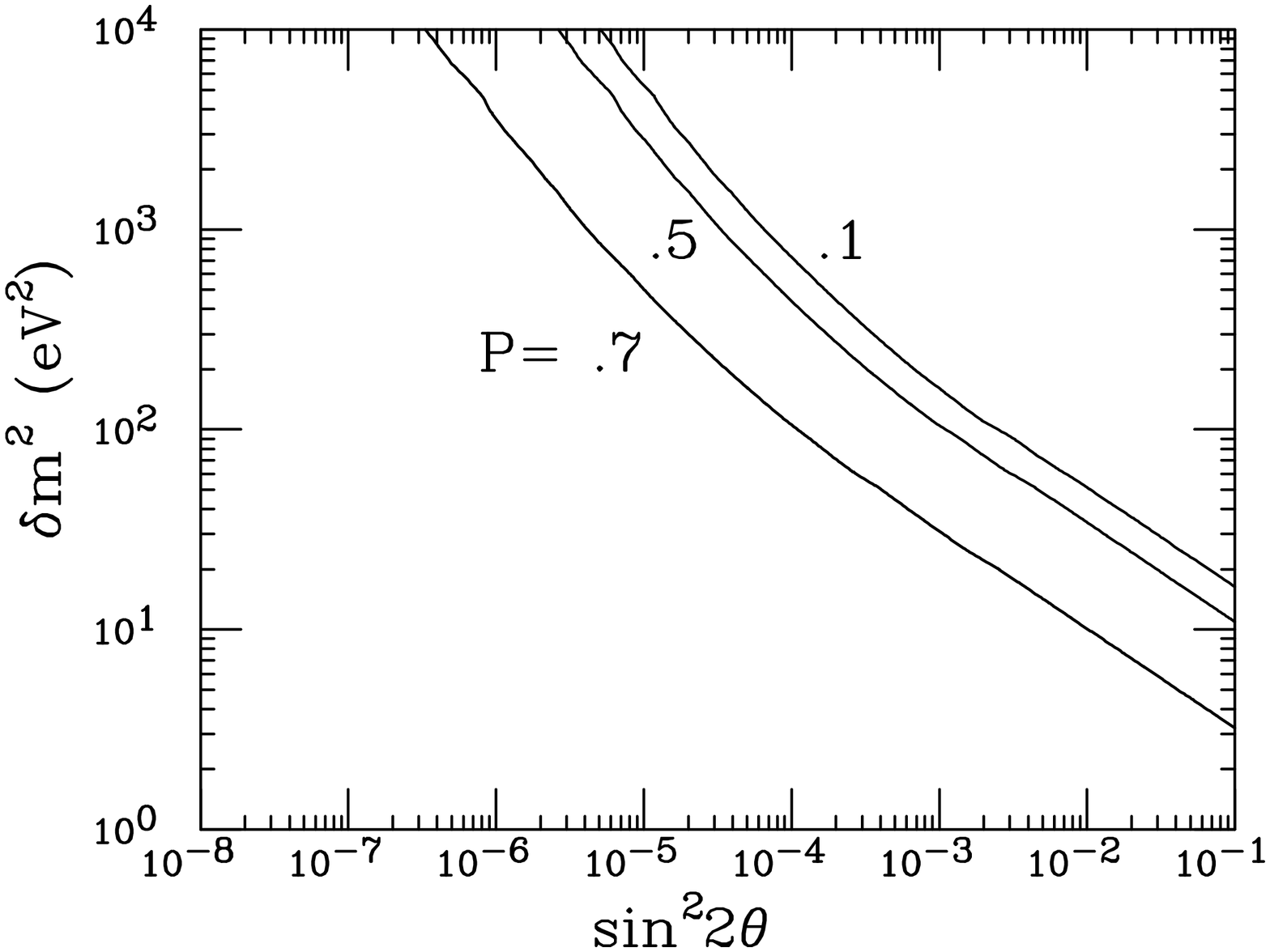,height=4.2cm,width=6.0cm,angle=90}
}
\vskip -4.22cm
{\hglue 5.4cm
\psfig{file=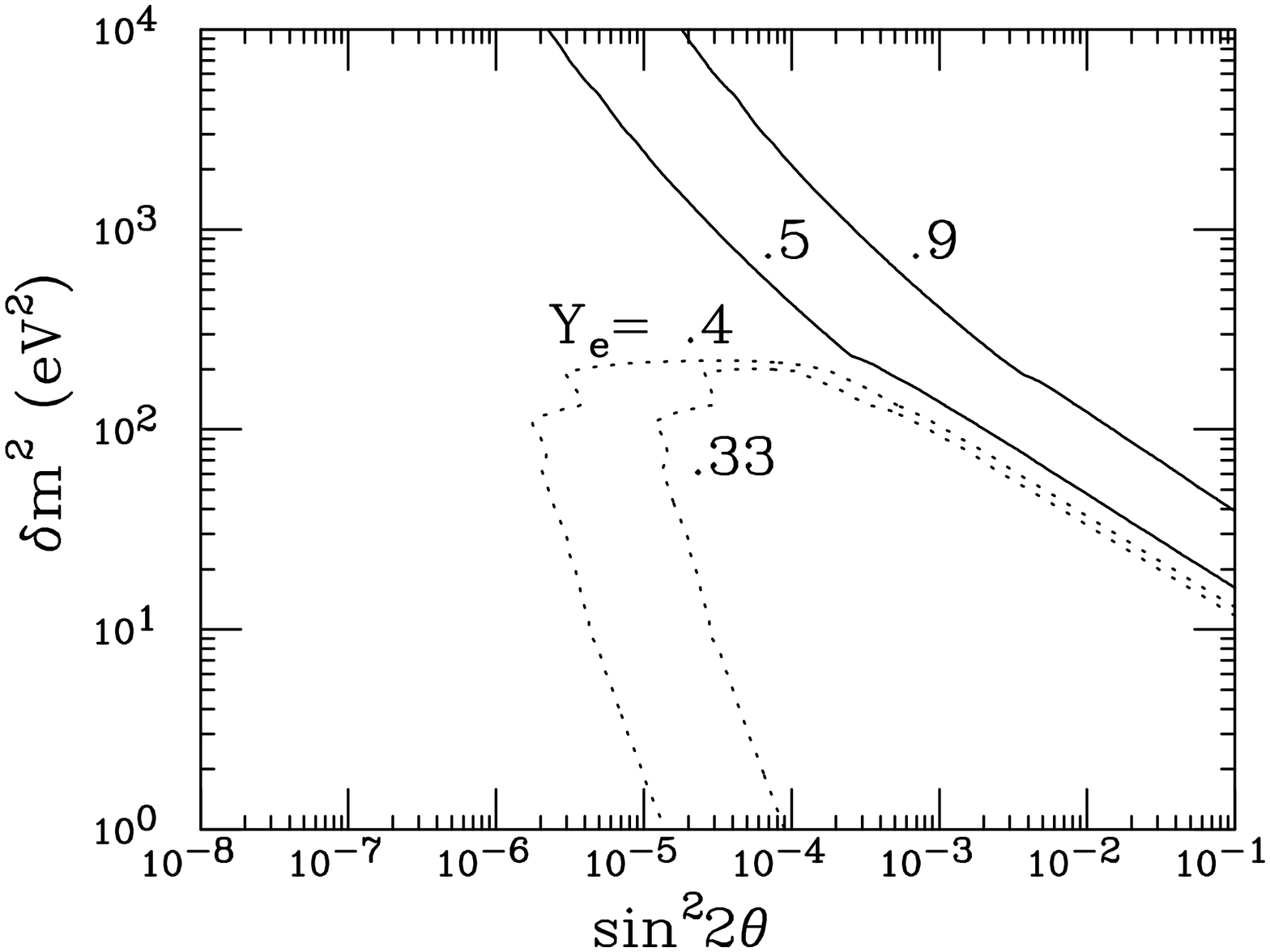,height=4.2cm,width=6.0cm,angle=90}
}
\vglue -0.63cm
\caption{Upper panel shows the contour plot of 
the reheating rate $R$, which is normalized to 1 
in the absence of oscillation, in 
$\delta m^2 - \sin^2 2\theta$ plane. 
Lower left palen shows the contour plot of the survival 
probability $P$ for the $\bar\nu_e\rightarrow\bar\nu_s$ conversion. 
Lower right panel shows the contour plot for the electron 
concentration $Y_e$ in the region relevant for $r$-process. 
(From ref. \protect{\cite{sterile}}.)
}
\label{fig:sterile}
\vskip -0.5cm
\end{figure}

We present our results in Fig. \ref{fig:sterile}. 
We can conclude from the first plot (upper panel) 
that if the neutrino re-heating is essential 
for successful supernova explosion the parameter region 
right to the curve, say $R=0.5$, is disfavoured. 
From the second plot (lower left) we conclude that the successful 
observation of the $\bar\nu_e$ signal from supernova 
SN1987A in Kamiokande and IMB detectors \cite{kamimb} implies 
the absence of significant conversion 
of $\bar{\nu}_e\rightarrow\bar{\nu}_s$, 
disfavouring the parameter region right to the 
curve, say $P=0.5$. 
In the third plot (lower right panel)
in the region right to the curve $Y_e=0.5$ the value of $Y_e$ is 
larger than 0.5 and hence the $r$-process is forbidden
whereas in the region delimited by the curve $Y_e=0.4$ 
the value of $Y_e$ could be decreased compared to 
the standard case, leading to the enhancement of 
$r$-process. 
\vskip -0.2cm
\section{Conclusion}
We have discussed the impact of the resonant conversion 
induced by some non-standard properties of neutrinos.  
In particular we focussed on the conversion induced 
by neutrino transition magnetic moment, 
flavor non-universality, FCNC interaction in 
SUSY models with broken $R$ parity, and 
mixing with some sterile state. 
We have shown that some significant effects on 
supernova physics are expected  and in some case 
we can derive bounds on neutrino parameters from the shock 
re-heating, $r$-process as well as SN1987A $\bar{\nu}_e$ 
signal arguments.  
We note that these bounds are first of all supernova 
model dependent, but complementary to the ones we 
obtain in the laboratory experiments and sometimes 
they are happen to be more stringent. 
\section*{Acknowledgments}
The author has been supported by a DGICYT postdoctral fellowship 
and by grant PB95-1077 and by TMR network grant ERBFMRXCT960090. 
The author would like to thank Anna Rossi for reading the manuscript. 

\end{document}